\documentclass[iop]{emulateapj}
\usepackage{apjfonts}

\shorttitle{Optimal strategies for continuous wave detection}
\shortauthors{Ellis et. al.}

\usepackage{amsmath,amssymb}
\usepackage{graphicx}
\usepackage{subfigure}
\newcommand{\be}{\begin{equation}}
\newcommand{\ee}{\end{equation}}
\newcommand{\omhat}{\hat{\Omega}}
\newcommand{\phat}{\hat{p}}
\newcommand{\hplus}{h_+}
\newcommand{\hcross}{h_{\times}}

\newcommand{\lp}{\left(}
\newcommand{\rp}{\right)}
\newcommand{\bb}{\begin{bmatrix}}
\newcommand{\eb}{\end{bmatrix}}

\begin{document}

\title{Optimal strategies for continuous gravitational wave detection in pulsar timing arrays}

\author{Justin A. Ellis\altaffilmark{1}, Xavier Siemens\altaffilmark{1}, and Jolien D.E.  Creighton\altaffilmark{1}}

\altaffiltext{1}{Center for Gravitation and Cosmology, University of Wisconsin Milwaukee, Milwaukee WI, 53211}




\begin{abstract}
Supermassive black hole binaries (SMBHBs) are expected to emit continuous 
gravitational waves in the pulsar timing array (PTA) frequency band 
($10^{-9}$--$10^{-7}$ Hz). The development of data analysis techniques 
aimed at efficient detection and characterization of these signals is  
critical to the gravitational wave detection effort.
In this paper we leverage methods developed for LIGO continuous 
wave gravitational searches, and explore the use of the $\mathcal{F}$-statistic
for such searches in pulsar timing data. 
\citealt{bs12} have already used this approach in the context of PTAs
to show that one can resolve multiple SMBHB sources in the sky. 
Our work improves on several aspects of prior continuous wave 
search methods developed for PTA data analysis. The algorithm is implemented 
fully in the time domain, which naturally deals with the irregular sampling typical 
of PTA data and avoids spectral leakage problems associated with frequency domain methods. 
We take into account the 
fitting of the timing model, and have generalized our approach to 
deal with both correlated and uncorrelated colored noise sources. We  
also develop an incoherent detection statistic that maximizes over all 
pulsar dependent contributions to the likelihood. 
To test the 
effectiveness and sensitivity of our detection statistics, we perform 
a number of monte-carlo simulations. 
We produce sensitivity curves for PTAs of various configurations,
and outline an implementation of a fully functional data analysis pipeline.
Finally, we present a derivation of the likelihood maximized over the gravitational 
wave phases at the pulsar locations, which results in a vast reduction of the 
search parameter space.  
\end{abstract}

\maketitle

\section{Introduction}
In the next few years pulsar timing arrays (PTAs) are expected
to detect gravitational waves (GWs) in the frequency range
$10^{-9}$--$10^{-7}$ Hz.  Potential sources of GWs in this frequency range include
supermassive black hole binary systems (SMBHBs) \citep{svc08}, 
cosmic (super)strings \citep{Olmez:2010bi}, inflation \citep{sa79}, and 
a first order phase transition at the QCD scale \citep{ccd+10}.
The community has thus far mostly focused on stochastic backgrounds produced 
by these sources, but they can manifest themselves in different ways. Cosmic strings and 
SMBHBs (in highly eccentric orbits) can also produce GW 
bursts \citep{dv01,smc07,lss09} in which the duration of the GW signal is much 
less than the observation time. Sufficiently nearby 
single SMBHBs may produce detectable continuous waves with periods on the order of 
years \citep{wl03,svv09,sv10}. The concept of a PTA, an array of accurately timed 
millisecond pulsars, was first conceived of over two decades ago \citep{r89,fb90}. 
Twenty years later three main PTAs are in full operation around the world: the 
North American Nanohertz Observatory for Gravitational waves (NANOGrav; \cite{jfl+09}), 
the Parkes Pulsar Timing Array (PPTA; \cite{m08}),
and the European Pulsar Timing Array (EPTA; \cite{jsk+08}).
The three PTAs collaborate  
to form the International Pulsar Timing Array (IPTA; \cite{haa+10}). 

Prior to the establishment of PTAs, \citet{jll+04} 
used existing pulsar data to rule out the proposed SMBHB system 3C66B, a 
possible source of continuous GWs (the mass of the proposed system 
has since been been lowered 
significantly \citep{ios10} so that it not likely to be detectable 
with current PTAs). In this work, the authors looked for the signature 
of a continuous GW in real pulsar data through the use of Lomb-Scargle 
periodograms and suggested a method for directed searches of known 
sources. \citet{yhj+10} also relied on the Lomb-Scargle periodogram 
to determine the sensitivity of the PPTA to continuous GW sources as a 
function of GW frequency. \citet{vl10} developed a bayesian framework
aimed at the detection of GW memory in PTAs; however, the authors mention
that the methods presented could be used for continuous GW sources as well.
 \citet{sv10} use an Earth-term only signal model to perform a 
 study of SMBHB parameters measurable with PTAs using a Fisher matrix approach.
  \citet{cc10} have developed a Bayesian Markov 
Chain Monte-Carlo (MCMC) data analysis algorithm for parameter estimation of a 
SMBHB system in which the pulsar term is taken into account 
in the detection scheme, thereby increasing the SNR and improving the accuracy of
the GW source location on the sky. Recently, \citet{lwk+11} have developed parameter estimation 
techniques based on vector Ziv-Zakai bounds incorporating the pulsar 
term and have placed limits on the minimum detectable amplitude of a continuous 
GW source. In this work, the authors also propose a method of combining timing parallax
measurements with single-source GW detections to improve pulsar distance measurements.

In the context of searches for continuous gravitational waves from spinning 
neutron stars in LIGO, \citet{jks98} developed the so-called $\mathcal{F}$-statistic, 
the logarithm of the likelihood ratio maximized over some of the signal 
parameters. \citet{cs05} later generalized the $\mathcal{F}$-statistic to multi-detector
networks. Very recently, \citet{bs12} have used the $\mathcal{F}$-statistic 
to show that in PTA data multiple SMBHB sources can be resolved in the sky.  
In this paper we build on this work, and improve on a number aspects of 
prior continuous wave search methods developed for PTA data analysis. 

In Section \ref{sec:signalmodel} we review the signal model. In Section \ref{sec:Fstat} we
discuss the $\mathcal{F}$-statistic in the context of PTA data. Unlike LIGO 
implementations of the $\mathcal{F}$-statistic, our algorithm is implemented 
fully in the time domain. This naturally deals with the irregular sampling of 
PTA data and avoids the spectral leakage problems that arise when frequency 
domain methods are used on such data.
We also account for the timing model: fitting out pulsar parameters removes signal power
at low frequencies, at frequencies near $1$~yr$^{-1}$ and $2$~yr$^{-1}$ due to 
sky location, proper motion, and parallax fitting, and for pulsars in binaries, at 
frequencies near the binary orbital frequency.
Our approach also naturally incorporates colored noise sources, both uncorrelated 
and correlated (for the case when the dominant noise source is a gravitational 
wave stochastic background). We also develop an incoherent detection statistic 
that maximizes over all pulsar dependent contributions to the likelihood. 
To test the effectiveness and sensitivity of our detection statistics, 
in Section \ref{sec:implementation} we perform a number of monte-carlo 
simulations. We produce sensitivity curves for PTAs of various configurations, 
and show that the performance of the incoherent statistic is comparable to the 
coherent $\mathcal{F}$-statistic. We also present an outline of the implementation 
of a continuous wave search pipeline. Finally, in Section \ref{sec:summary} we 
summarize our results
and conclude with a derivation of the 
likelihood maximized over the gravitational wave phases at the pulsar locations, 
which results in a vast reduction of the search parameter space. We leave the 
exploration of this new statistic for future work.

\section{The Signal Model}
\label{sec:signalmodel}

In this section we will briefly review the form of the residuals induced by 
a non-spinning SMBHB in a circular orbit and introduce our notation. The GW 
is defined as a metric perturbation to flat space time defined as
\be
h_{ab}(t,\omhat)=e_{ab}^+(\omhat)\hplus(t,\omhat)+e_{ab}^{\times}(\omhat)\hcross(t,\omhat),
\ee
where $\omhat$ is the unit vector pointing from the GW source to the 
SSB, $\hplus$, $\hcross$ and $e_{ab}^A$ ($A=+, \times$) are the  
polarization amplitudes and polarization tensors, respectively. The 
polarization tensors can be converted to the Solar System Barycenter 
(SSB) by the following transformation. Following \cite{w87} we write
\begin{align}
e_{ab}^+(\omhat)&=\hat{m}_a\hat{m}_b-\hat{n}_a\hat{n}_b,\\
e_{ab}^{\times} (\omhat)&=\hat{m}_a\hat{n}_b+\hat{n}_a\hat{m}_b,
\end{align}
where
\begin{align}
\omhat &=-(\sin\theta\cos\phi)\hat{x}-(\sin\theta\sin\phi)\hat{y}-(\cos\theta)\hat{z},\\
\hat{m} &=-(\sin\phi)\hat{x}+(\cos\phi)\hat{y},\\
\hat{n} &=-(\cos\theta\cos\phi)\hat{x}-(\cos\theta\sin\phi)\hat{y}+(\sin\theta)\hat{z}.
\end{align}
In this coordinate system, $\theta=\pi/2-\delta$ and $\phi=\alpha$ are the polar and azimuthal angles of the source, respectively, where $\delta$ 
and $\alpha$ are declination and right ascension in usual celestial coordinates. 

We will write our GW induced pulsar timing residuals in the following form:
\be
s(t,\omhat)=F^{+}(\omhat)\Delta s_{+}(t)+F^{\times}(\omhat)\Delta s_{\times}(t),
\ee
where 
\be
\Delta s_{A}(t)=s_{A}(t_{p})-s_{A}(t_{e}),
\ee
and $t_{e}$ and $t_{p}$ are the times at which the GW passes the Earth and pulsar, respectively, and
the index $A$ labels polarizations. 
The functions $F^{A}(\omhat)$ are known as antenna pattern functions and are defined by
\begin{align}
F^{+}(\omhat)&=\frac{1}{2}\frac{(\hat{m}\cdot\phat)^{2}-(\hat{n}\cdot\phat)^{2}}{1+\omhat\cdot\phat}\\
F^{\times}(\omhat)&=\frac{(\hat{m}\cdot\phat)(\hat{n}\cdot\phat)}{1+\omhat\cdot\phat},
\end{align}
where $\phat$ is the unit vector pointing from the Earth to the pulsar. Also, from geometry we can write
\be
\label{eq:pTime}
t_{p}=t_{e}-L(1+\omhat\cdot\phat).
\ee
Given these definitions, we can write the GW contributions to the timing residuals as \citep{w87,cc10}
\begin{align}
\begin{split}
\label{eq:rplus}
s_{+}(t)&=\frac{\mathcal{M}^{5/3}}{D\omega(t)^{1/3}}\Big[-\sin[2(\Phi(t)-\Phi_{0})](1+\cos^{2}\iota)\cos2\psi\\
&-2\cos[2(\Phi(t)-\Phi_{0})]\cos\iota\sin2\psi\Big]
\end{split}\\
\begin{split}
\label{eq:rcross}
s_{\times}(t)&=\frac{\mathcal{M}^{5/3}}{D\omega(t)^{1/3}}\Big[-\sin[2(\Phi(t)-\Phi_{0})](1+\cos^{2}\iota)\sin2\psi\\
&+2\cos[2(\Phi(t)-\Phi_{0})]\cos\iota\cos2\psi\Big],
\end{split}
\end{align}
where
\be
\Phi(t)=\frac{1}{32\mathcal{M}^{5/3}}\lp\omega_{0}^{-5/3}-\omega(t)^{-5/3}\rp
\ee
and
\be
\label{eq:worb}
\omega(t)=\lp \omega_{0}^{-8/3}-\frac{256}{5}\mathcal{M}^{5/3}t \rp^{-3/8}.
\ee
For reasons that will become clear later, we write the residuals for pulsar $\alpha$ in the following form
\be
\begin{split}
\label{eq:resids}
{r}_{\alpha}(t,\omhat)&=s_{\alpha}(t,\omhat)+n_{\alpha}(t)\\
&=\sum_{i=1}^{4}\left[a_{i}(\zeta,\iota,\Phi_{0},\psi){A}_{\alpha}^{i}(t,\theta,\phi,\omega_{0})\right]\\
&+p_{\alpha}(t,\zeta,\iota,\Phi_{0},\psi,\theta,\phi,\omega_{0},L_{\alpha})+n_{\alpha}(t),
\end{split}
\ee
where $\zeta=\mathcal{M}^{5/3}D^{-1}$, $n_{\alpha}(t)$ is the noise in each pulsar and
\be
p_{\alpha}=F^{+}(\omhat) s_{+}(t_{p})+F^{\times}(\omhat)s_{\times}(t_{p}).
\ee
Hereon we will refer to the summation term as the Earth term 
and $p$ as the pulsar term. We write the combination of chirp mass and distance 
to the binary as one parameter because the two can not be disentangled 
unless there is a measurement of $\dot{f}$, which we do not consider here. 
It is customary to label the parameters $(\zeta,\iota,\Phi_{0},\psi)$ and 
$(\theta,\phi,\omega_{0})$ extrinsic and intrinsic parameters \citep{jks98}, 
respectively.  We then define the amplitudes and time dependent basis functions 
\be
\label{eq:amps}
\begin{split}
a_{1}&=\zeta\left[ (1+\cos^{2}\iota)\cos\Phi_{0}\cos2\psi +2\cos\iota\sin\Phi_{0}\sin2\psi \right]\\
a_{2}&=-\zeta\left[ (1+\cos^{2}\iota)\sin\Phi_{0}\cos2\psi -2\cos\iota\cos\Phi_{0}\sin2\psi \right]\\
a_{3}&=\zeta\left[ (1+\cos^{2}\iota)\cos\Phi_{0}\sin2\psi -2\cos\iota\sin\Phi_{0}\cos2\psi \right]\\
a_{4}&=-\zeta\left[ (1+\cos^{2}\iota)\sin\Phi_{0}\sin2\psi +2\cos\iota\cos\Phi_{0}\cos2\psi \right]
\end{split}
\ee
and
\be
\label{eq:basis}
\begin{split}
{A}_{\alpha}^{1}&={F_{\alpha}^{+}(\omhat)}{\omega(t)^{-1/3}}\sin(2\Phi(t))\\
{A}_{\alpha}^{2}&={F_{\alpha}^{+}(\omhat)}{\omega(t)^{-1/3}}\cos(2\Phi(t))\\
{A}_{\alpha}^{3}&={F_{\alpha}^{\times}(\omhat)}{\omega(t)^{-1/3}}\sin(2\Phi(t))\\
{A}_{\alpha}^{4}&={F_{\alpha}^{\times}(\omhat)}{\omega(t)^{-1/3}}\cos(2\Phi(t)).
\end{split}
\ee

Throughout this work we assume that the source is slowly evolving (i.e. the phase is independent of the chirp mass) and $\omega(t)\approx\omega_{0}$ and $\Phi(t)\approx \omega_{0}t$.

\section{The Likelihood Function and the $\mathcal{F}$-statistic }
\label{sec:Fstat}

Here we will introduce our formalism and derive the likelihood and 
$\mathcal{F}$-statistic (the likelihood maximized over extrinsic parameters) 
for PTAs. We will also discuss the statistics of the $\mathcal{F}$-statistic 
in the presence and absence of a signal and show that we obtain the expected 
behavior for PTA data.

\subsection{Likelihood}
For a pulsar timing array with $M$ pulsars we define the likelihood function of 
the noise as multivariate gaussian
\be
p(\mathbf{n})=\frac{1}{\sqrt{2\pi\mathbf{\Sigma}_{n}}}\exp\lp -\frac{1}{2}\mathbf{n}^{T}
\mathbf{\Sigma}_{n}^{-1}\mathbf{n} \rp,
\ee
where
\be
\mathbf{n}=\bb {n}_{1} \\ {n}_{2}\\ \vdots \\ {n}_{M} \eb
\ee 
is the vector of the noise time-series for all pulsars,
\be
\label{eq:cov}
\mathbf{\Sigma}_{n}=\bb  \Sigma_{n,1} & S_{12} & \hdots & S_{1M}\\ 
S_{21} & \Sigma_{n,2} & \hdots & S_{2M}\\
\vdots & \vdots & \ddots & \vdots\\
S_{1M} & S_{2M} & \hdots & \Sigma_{n,M}\eb
\ee
is the multivariate covariance matrix, and 
\begin{align}
\Sigma_{n,i}&=\langle \mathbf{n}_{i}\mathbf{n}_{i}\rangle\\
S_{ij}&=\langle \mathbf{n}_{i}\mathbf{n}_{j}\rangle\big|_{i\ne j}
\end{align}
\begin{figure}[t]
  \begin{center}
	\includegraphics[width=0.49\textwidth]{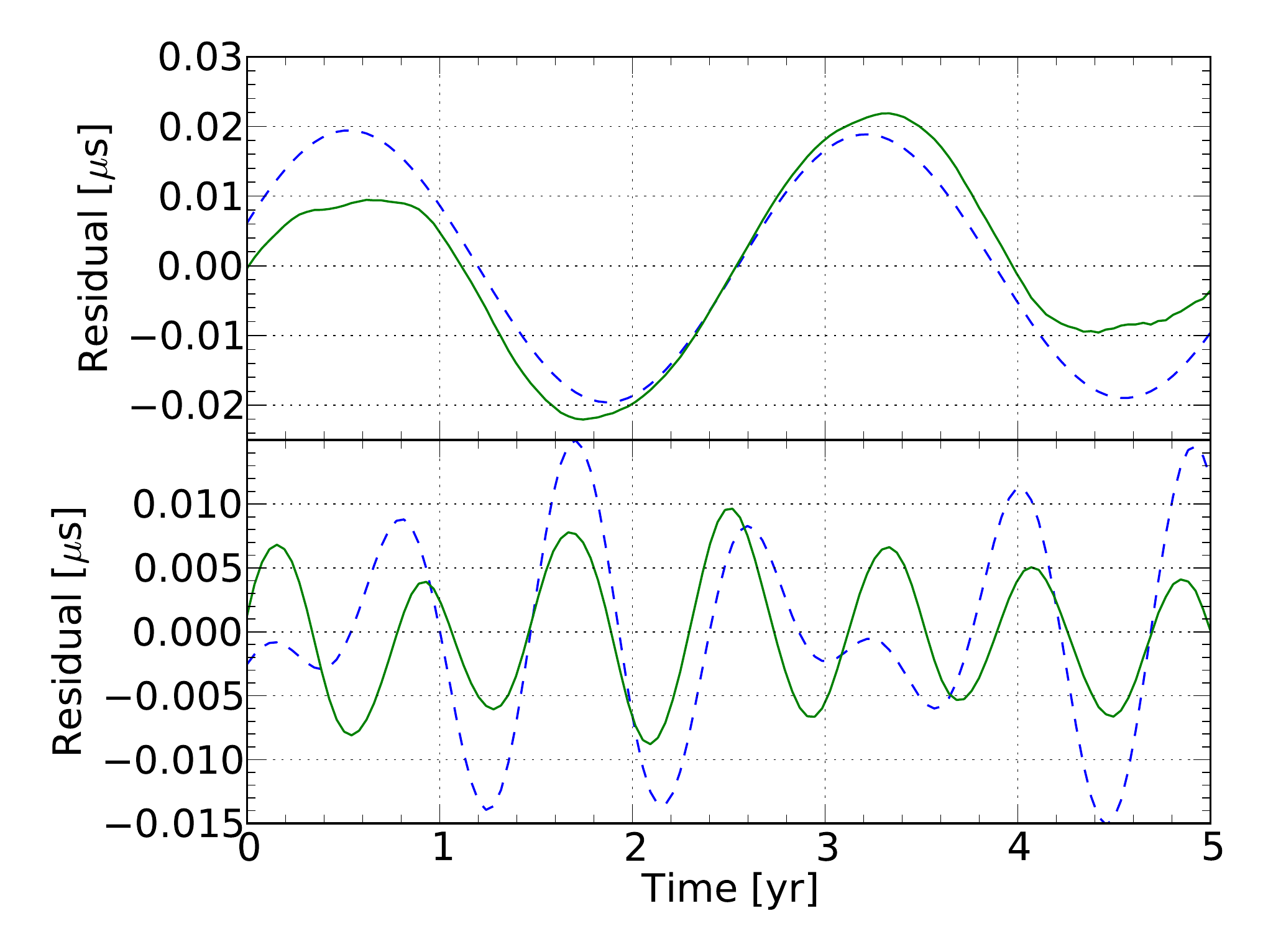}
  \end{center}
  \caption{SMBHB waveforms in two different regimes. Each plot shows 
the waveform before (dotted blue) and after fitting (solid green) 
for a full timing model including spin-down, astrometeric and binary 
parameters. Top Panel: The Earth and pulsar terms at the same frequency. 
Bottom Panel: The Earth term and pulsar term at different frequencies.}
\label{fig:wave}
\end{figure}
are the auto-covariance and 
cross-covariance matrices of the pulsar noise, respectively. It is important 
to note that in the case 
of uncorrelated noise, the off-diagonal cross covariance matrices vanish.  
In order to time pulsars, a timing model is fitted out of the pulsar 
times-of-arrival (TOAs) via a weighted least squares fitting routine 
\citep{hem06}. This procedure can be expressed via a data-independent linear 
operator $\mathbf{R}$ (see \citealt{dfg+12} for details) so that
\be
\tilde{\mathbf{n}}=\mathbf{R}\mathbf{n},
\ee
where
\be
\mathbf{R}=\bb {R}_{1} \\{R}_{2}\\ \vdots \\{R}_{M} \eb
\ee
is a vector of matrices $R_i$, the fitting operators for each pulsar, 
and $\tilde{\mathbf{n}}$ is the post-fit noise. We can 
see the effect of this fitting procedure on the waveforms in 
Fig. \ref{fig:wave} where the waveform is changed, quite significantly, 
from its pre-fit form. It is straightforward to show that the 
likelihood for the fitted $\tilde{\mathbf{n}}$ is
\be
p(\tilde{\mathbf{n}})=\frac{1}{\sqrt{2\pi\mathbf{\Sigma}_{\tilde{n}}}}\exp\lp 
-\frac{1}{2}\tilde{\mathbf{n}}^{T}\mathbf{\Sigma}_{\tilde{n}}^{-1}\tilde{\mathbf{n}} \rp,
\ee
where 
\be
\mathbf{\Sigma}_{\tilde{n}}=\langle \tilde{\mathbf{n}}\tilde{\mathbf{n}}^{T} 
\rangle=\mathbf{R}\langle \mathbf{n}\mathbf{n}^{T}\rangle \mathbf{R}^{T}.
\ee
The fitted residuals can therefore be written as
\be
\mathbf{r}=\mathbf{R}\lp \mathbf{s}+\mathbf{n} \rp= \tilde{\mathbf{s}}
+\tilde{\mathbf{n}},
\ee
where 
\be
\mathbf{r}=\bb {r}_{1} \\ {r}_{2}\\ \vdots \\ {r}_{M} \eb,\hspace{2cm} 
\mathbf{s}=\bb {s}_{1} \\ {s}_{2}\\ \vdots \\ {s}_{M} \eb
\ee
are the residual data and signal template for each pulsar, respectively. 
We can therefore write the likelihood of the data $\mathbf{r}$ given some 
signal template $\mathbf{s}$
 \be
p(\mathbf{r}|\mathbf{s})=\frac{1}{\sqrt{2\pi\mathbf{\Sigma}_{\tilde{n}}}}\exp\lp 
-\frac{1}{2}(\mathbf{r}-\tilde{\mathbf{s}})^{T}\mathbf{\Sigma}_{\tilde{n}}^{-1}
(\mathbf{r}-\tilde{\mathbf{s}}) \rp.
 \ee

We define the inner product for two time vectors $\mathbf{x}$ and $\mathbf{y}$ 
using the noise covariance matrix $\mathbf{\Sigma}_{n}$ as
 \be
 \label{eq:inner}
 \lp \mathbf{x}|\mathbf{y} \rp=\mathbf{x}^{T}\mathbf{\Sigma}_{\tilde n}^{-1}\mathbf{y}.
 \ee
In this notation we can write the log of the likelihood ratio as
 \be
 \label{eq:llike}
 \ln \, \Lambda =\ln \,\frac{p(\mathbf{r}|\mathbf{s})}{p(\mathbf{r}|0)}
=(\mathbf{r}|\tilde{\mathbf{s}})-\frac{1}{2}(\tilde{\mathbf{s}}|\tilde{\mathbf{s}}).
 \ee
It is worth pointing out that finding the inverse of $\mathbf{\Sigma}_{\tilde n}$ is computationally 
intensive. Aside from it being a very large matrix, the fitting procedure results in 
loss of degrees of freedom in the data which makes $\mathbf{\Sigma}_{\tilde n}$ singular.
Inverting this matrix therefore requires singular value decomposition. 

In most realistic scenarios we can assume that 
the off-diagonal cross-covariance matrices are small and expand the inverse 
of Eq. \ref{eq:cov} in a Neumann series (see Eq. 72 of \citealt{abc+08} for 
details).  In the simulations shown later in the paper we will assume that any 
correlated noise is much 
less than the uncorrelated part, thus we treat $\mathbf{\Sigma}_{\tilde n}$ as a 
block diagonal matrix of the auto-covariance matrices for each pulsar.

\subsection{The Earth-term $\mathcal{F}$-statistic}
\label{sec:F}
\begin{figure*}[t]
  \begin{center}
  \subfigure[]{
  	\includegraphics[width=0.48\textwidth]{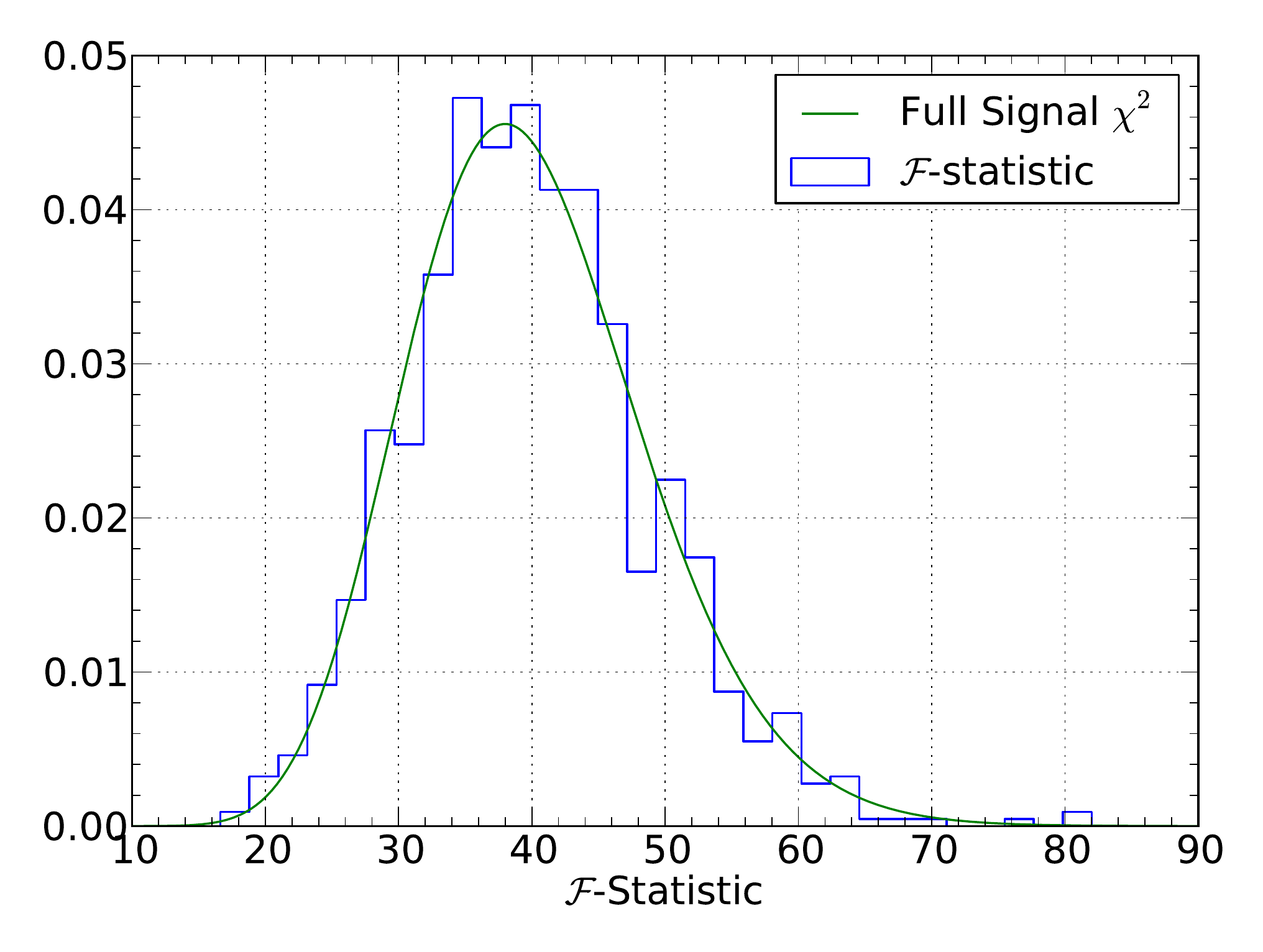}
	\label{fig:pdf_a}}
  \subfigure[]{
  	\includegraphics[width=0.48\textwidth]{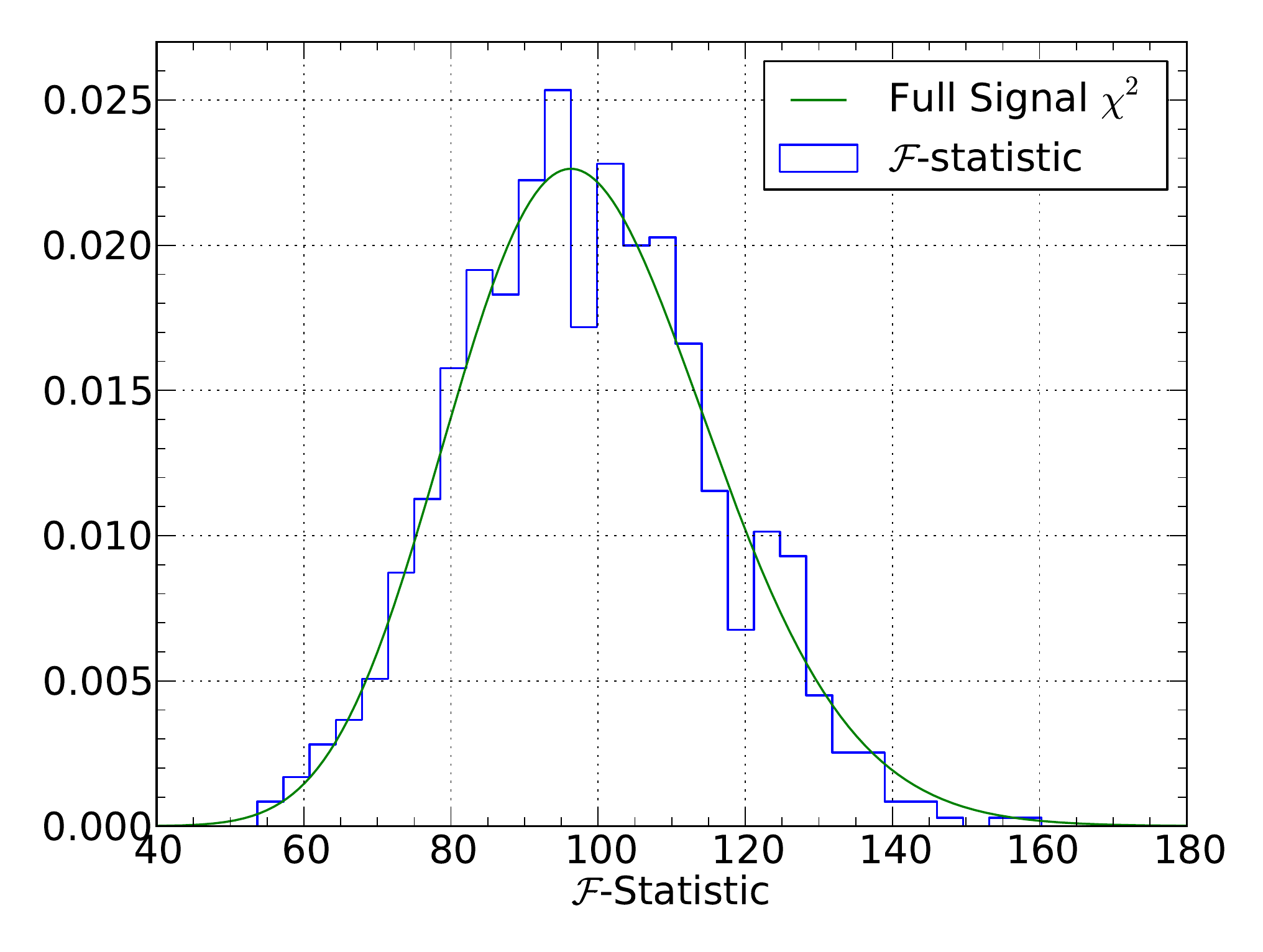}
	\label{fig:pdf_b}}\\
  \subfigure[]{
  	\includegraphics[width=0.48\textwidth]{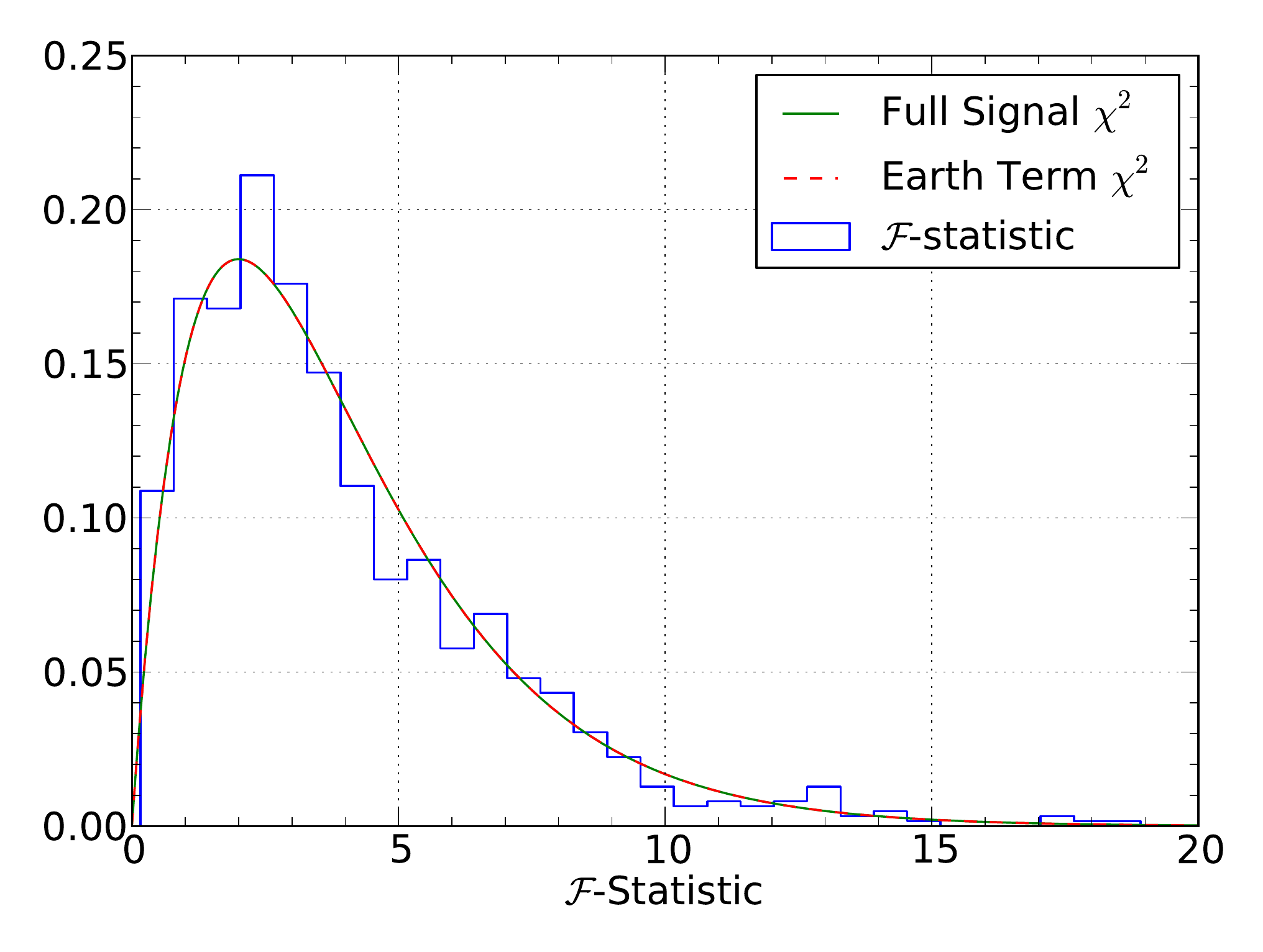}
	\label{fig:pdf_c}}
  \subfigure[]{
  	\includegraphics[width=0.48\textwidth]{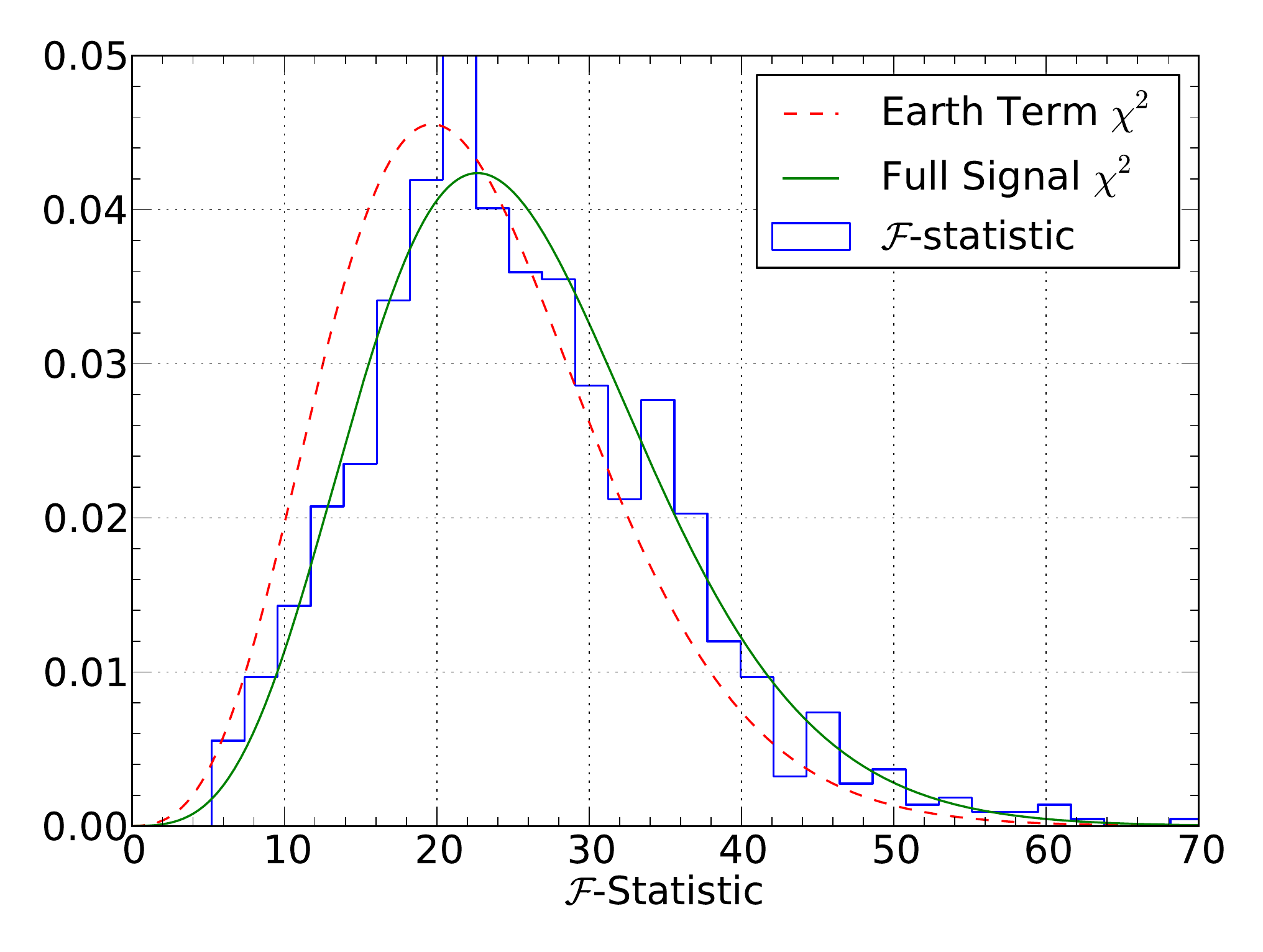}
  	\label{fig:pdf_d}}
   \end{center}
  \caption{Histograms and expected probability distribution functions of $2\mathcal{F}_{p}$ and $2\mathcal{F}_{e}$ in the absence and presence of a signal for 20 pulsars. Each simulation was done with the search parameters fixed and 1000 realizations of white gaussian noise. (a): distribution of $2\mathcal{F}_{p}$ in the absence of a signal. (b): distribution of $2\mathcal{F}_{p}$ in the presence of a signal with non-centrality parameter $\rho^{2}$. (c): distribution of $2\mathcal{F}_{e}$ in the absence of a signal. (d): distribution of $2\mathcal{F}_{e}$ in the presence of a signal.  The dashed (red) and solid (green) curves have the same meaning as in Fig. \ref{fig:dropPterm}.}
\label{fig:pdf}
\end{figure*}

We now analytically maximize over the extrinsic 
parameters $(\zeta,\iota,\Phi_{0},\psi)$ in the signal model. 
A very similar calculation was first done by \cite{jks98} in the 
context of LIGO, subsequently by \cite{cp07} in the 
context of LISA, and very recently by \cite{bs12} in the context 
of pulsar timing. For clarity, here we review this calculation in the 
notation introduced above. For this calculation we treat the pulsar 
term  as a noise source and write  our signal template in 
the form
\be
\mathbf{s}(t,\omhat)=\sum_{i=1}^{4}a_{i}(\zeta,\iota,\Phi_{0},\psi)\mathbf{A}^{i}(t,\theta,\phi,\omega_{0}),
\ee
where
\be
\mathbf{A}^{i}=\bb {A}^{i}_{1} \\{A}^{i}_{2}\\ \vdots \\{A}^{i}_{M} \eb.
\ee
Later we will explain the circumstances under which it is safe to drop the 
pulsar term. We can now write the log-likelihood as
\be
\ln\,\Lambda=a_{i}(\mathbf{r}|\mathbf{A}^{i})-\frac{1}{2}(\mathbf{A}^{i}|\mathbf{A}^{j})a_{i}a_{j}=a_{i}\mathbf{N}^{i}-\frac{1}{2}\mathbf{M}^{ij}a_{i}a_{j}.
\ee
Maximizing the log-likelihood ratio over the four amplitudes $a_{i}$ gives
\be
\begin{split}
\label{eq:maxLike}
\frac{\partial \ln\,\Lambda}{\partial a_{k}}&=0=\mathbf{N}^{i}\delta^{k}_{i}-\frac{1}{2}\mathbf{M}^{ij}a_{j}\delta^{k}_{i}-\frac{1}{2}\mathbf{M}^{ij}a_{i}\delta^{k}_{j}\\
&=\mathbf{N}^{k}-\mathbf{M}^{ik}a_{i},
\end{split}
\ee
yielding the maximum likelihood estimators for the four amplitudes
\be
a_{i}=\mathbf{M}_{ij}\mathbf{N}^{j},
\ee
where $\mathbf{M}_{ij}=(\mathbf{M}^{ij})^{-1}$. Substituting these back into the likelihood 
results in the $\mathcal{F}_{e}$-statistic
\be
2\mathcal{F}_{e}=\mathbf{N}^{i}\mathbf{M}_{ij}\mathbf{N}^{j}.
\label{eq:fstat22}
\ee
The statistics of $2\mathcal{F}_{e}$ are a $\chi^{2}$ with 4 degrees of freedom and a 
non-centrality parameter $\bar{\rho}^{2}$. It is straightforward to show that the 
expectation value is 
\be
\begin{split}
\label{eq:expFe}
\langle 2\mathcal{F}_{e}\rangle&=4+\bar{\rho}^{2}\\
&=4+(\tilde{\mathbf{s}}|\tilde{\mathbf{s}})+2(\tilde{\mathbf{p}}|\tilde{\mathbf{s}})+(\tilde{\mathbf{p}}|\mathbf{A}^{i})\mathbf{M}_{ij}(\tilde{\mathbf{p}}|\mathbf{A}^{j}),
\end{split}
\ee
where $\mathbf{p}$ is the functional form of the pulsar term and the second two terms 
in $\bar{\rho}^{2}$ are due to the fact that we have only included the Earth term in 
our templates $\mathbf{s}$. In Figs. \ref{fig:pdf_c} and \ref{fig:pdf_d} we can see 
that the probability distribution functions of $2\mathcal{F}_{e}$ follow the expected 
distributions in the absence and presence of a signal. While only the intrinsic 
parameters are formally searched over, it is also possible to get estimates of the 
maximized extrinsic parameters by constructing the following quantities \citep{cp07}:
\be
\begin{split}
A_{+}&=\sqrt{(a_{1}+a_{4})^{2}+(a_{2}-a_{3})^{2}}\\
&+\sqrt{(a_{1}-a_{4})^{2}+(a_{2}+a_{3})^{2}},
\end{split} 
\ee
\be
\begin{split}
A_{\times}&=\sqrt{(a_{1}+a_{4})^{2}+(a_{2}-a_{3})^{2}}\\
&-\sqrt{(a_{1}-a_{4})^{2}+(a_{2}+a_{3})^{2}}
\end{split}
\ee
and
\be
A=A_{+}+\sqrt{A_{+}^{2}+A_{\times}^{2}}.
\ee
It is then possible to recover the maximized parameters
\begin{align}
\iota&=\cos^{-1}\left( \frac{-A_{\times}}{A} \right),\label{eq:iota}\\
\psi&=\frac{1}{2}\tan^{-1}\left( \frac{A_{+}a_{4}-A_{\times}A_{1}}
{A_{\times}a_{3}+A_{+}a_{2}} \right),\label{eq:psi}\\
\Phi_{0}&=-\tan^{-1}\left( \frac{-(A_{\times}a_{1}-A_{+}a_{4})}
{(A_{+}a_{3}+A_{\times}a_{2})} \right),\label{eq:phi}\\
\zeta&=\frac{Ac}{4}\label{eq:zeta},
\end{align}
where $c=sgn({\sin2\psi})$. It is interesting to examine the case of one pulsar. 
In this case, Eq. \ref{eq:maxLike} has no solution because the matrix 
$\mathbf{M}$ is singular. The reason for this is that it is incorrect to 
write the residuals in the form of Eq. \ref{eq:resids} with four degrees 
of freedom. For one pulsar, the signal has only \emph{two} degrees of 
freedom: an amplitude and a phase, or equivalently, two unknown amplitudes, 
thereby making the maximization over \emph{four} independent amplitudes an 
ill-posed problem. Thus, at least two pulsars are needed to solve Eq. 
\ref{eq:maxLike}. It should be noted that it is straightforward to 
generalize this statistic to $N$ GW sources, we will simply have $4N$ 
independent amplitudes instead of just 4 (see \citealt{bs12} for more 
details). However, for simplicity in this work we will deal with just 
one GW source.

\subsubsection{Justification for dropping the pulsar term}

There are two cases 
in which the pulsar term is truly negligible to the 
$\mathcal{F}_{e}$-statistic and can be dropped from the analysis 
with no change in the statistics. 

\begin{figure*}[t]
  \begin{center}
  \subfigure[]{
  	\includegraphics[width=0.48\textwidth]{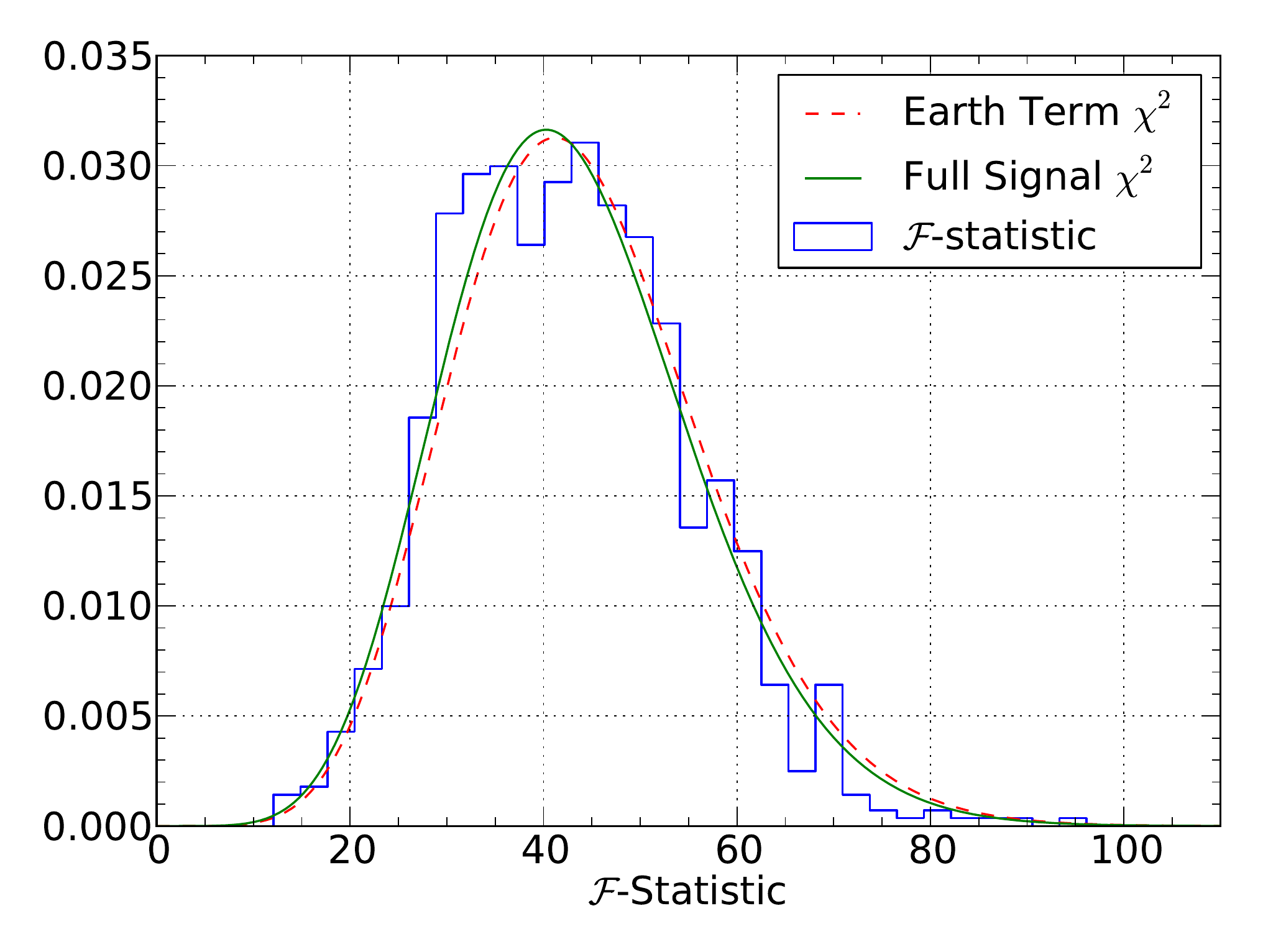}
	\label{fig:dropPterm_a}}
  \subfigure[]{
  	\includegraphics[width=0.48\textwidth]{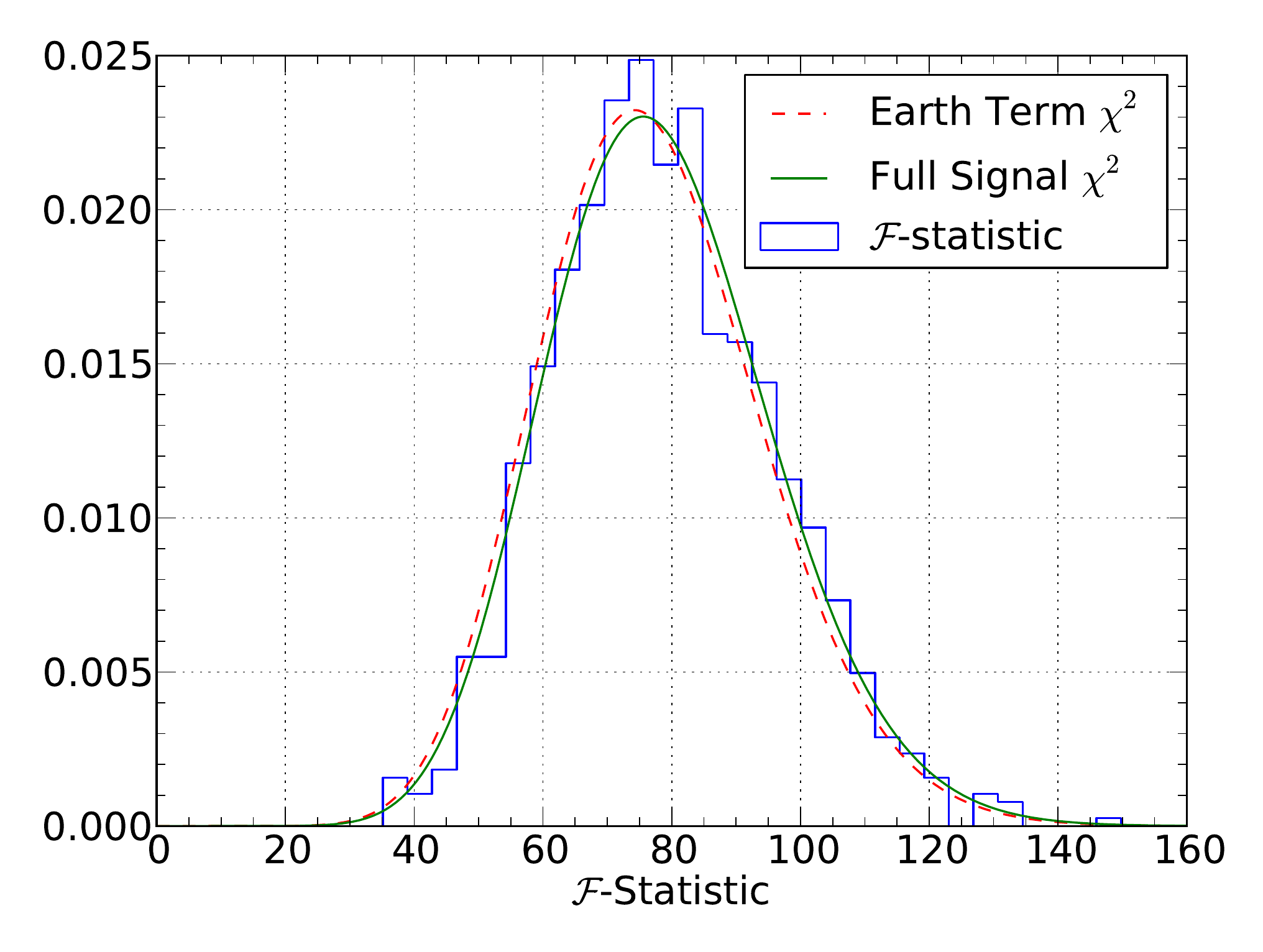}
	\label{fig:dropPterm_b}}
   \end{center}
  \caption{Probability distribution functions for $2\mathcal{F}_{e}$ 
in the limits that the pulsar term is negligible. (a): probability 
distribution function in the limit that all pulsar terms lie outside 
the Earth term frequency bin. (b): probability distribution function 
in the limit of large $M$ for overlapping Earth and pulsar term 
frequencies. The dashed (red) curve is a $\chi^{2}$ distribution 
with a non-centrality parameter assuming that only the Earth 
term is present in the data. The solid (green) curve is a 
$\chi^{2}$ distribution with non-centrality parameter 
$\bar{\rho}^{2}$ that takes both the Earth and pulsar term 
into account.}
\label{fig:dropPterm}
\end{figure*}

The first is the astrophysically 
likely scenario in which the evolution of the GW frequency is such 
that the Earth and pulsar terms are in different frequency bins 
(see e.g. Figure 2 of \citealt{sv10}). 
At the frequency of the Earth term the signal will build up coherently.
The pulsar term signals, even if they all happen to be at the same 
frequency, will not because they have different phases that depend 
on the the pulsar distances. This effect is illustrated in 
Fig. \ref{fig:dropPterm_a} where the reference $\chi^{2}$ 
distribution has 4 degrees of freedom 
and non-centrality parameter $(\tilde{\mathbf{s}}|\tilde{\mathbf{s}})$.

The second case is the less astrophysically likely scenario in 
which the Earth and pulsar term lie in the same frequency bin. In this 
case, there is still a phase difference between the Earth and pulsar terms. 
We expect that for a large number of pulsars the pulsar term signals will 
cancel because they all have different phases.
We can see from Fig. \ref{fig:pdf_d} that for a moderate number of pulsars 
($M=20$ in this case) the pulsar phases do not completely cancel and our 
measured values of the $\mathcal{F}_{e}$-statistic are higher than 
expected with just the earth term because the last two terms of Eq. 
\ref{eq:expFe} do not sum to zero. However, in the case of large $M$ 
($M\gtrsim50$) the pulsar term contributions sum approximately to zero, and 
again we have a $\chi^{2}$ distribution with 4 degrees of 
freedom and non-centrality parameter $(\tilde{\mathbf{s}}|\tilde{\mathbf{s}})$.

If we happen to detect a signal that falls into the intermediate 
category mentioned above where $M<50$ and some or all of the 
pulsar terms are in the same frequency bin as the Earth term, 
then this will create a bias in the recovered sky location but 
not in our ability to confidently detect the signal 
(see Fig.1 of \cite{ejm12}). This is 
because our detection criterion is the false alarm probability. 
As will be discussed in detail in Section \ref{sec:prob}, the 
false alarm probability only depends on the probability 
distribution function when the signal is \emph{absent}, 
and we can see from Fig. \ref{fig:pdf_a} that 
$2\mathcal{F}_{e}$ follows the expected distribution, 
because it is independent of the signal properties.

\subsection{The incoherent $\mathcal{F}$-statistic}
\label{sec:Fp}
It is indeed possible to include the pulsar term in our analysis 
if we operate in the low frequency (or low chirp mass) regime 
where the frequency evolution of the source is slow enough that 
the frequency at the Earth and the pulsar are essentially the 
same so that the signal is a sum of two sinusoids of different phases: 
the pulsar term and the Earth term. To understand this more quantitatively, 
consider the Taylor series expansion of the orbital frequency of 
Eq. \ref{eq:worb} evaluated at the pulsar time
\be
\label{eq:freqExpand}
\begin{split}
\omega(t_{p})&=\omega_{0}\lp 1-\frac{256}{5}\mathcal{M}^{5/3}\omega_{0}^{8/3}t_{p} \rp^{-3/8}\\
&\approx \omega_{0}\lp 1+\frac{96}{5}\mathcal{M}^{5/3}\omega_{0}^{8/3}\left[ t_{e}-L(1+\omhat\cdot\phat)\right] \rp.
\end{split}
\ee
From this, we can see that $\omega(t_{p})\approx\omega_{0}$ when
\be
\omega_{0}\ll \lp \frac{5}{96}\mathcal{M}^{-5/3}\left| \lp T-L(1+\omhat\cdot\phat)\rp\right| \rp^{3/8},
\ee
where $T$ is the total observation time. If we consider only one intrinsic parameter, $\omega_{0}$, then the template for pulsar $\alpha$ is
\be
s_{\alpha}(t,\omhat)=\sum_{i=1}^{2}b_{i\alpha}(\zeta,\iota,\psi,\Phi_{0},\phi_{\alpha},\theta,\phi)B_{\alpha}^{i}(t,\omega_{0}),
\ee 
where 
\be
\phi_{\alpha}=\omega L_{\alpha}(1+\omhat\cdot\phat_{\alpha})+\Phi_{0}
\ee
is the pulsar dependent phase. We can now write the pulsar dependent amplitudes and basis functions as
\begin{align}
\begin{split}
b_{1\alpha}&=\zeta\bigg[(1+\cos^{2}\iota)(F_{\alpha}^{+}\cos 2\psi+F_{\alpha}^{\times}\sin 2\psi)(\cos\Phi_{0}-\cos\phi_{\alpha})\\
&+ 2\cos\iota (F_{\alpha}^{+}\sin 2\psi-F_{\alpha}^{\times}\cos 2\psi)(\sin\Phi_{0}-\sin\phi_{\alpha})\bigg]
\end{split}\\
\begin{split}
b_{2\alpha}&=-\zeta\bigg[(1+\cos^{2}\iota)(F_{\alpha}^{+}\cos 2\psi+F_{\alpha}^{\times}\sin 2\psi)(\sin\Phi_{0}-\sin\phi_{\alpha})\\
&- 2\cos\iota (F_{\alpha}^{+}\sin 2\psi-F_{\alpha}^{\times}\cos 2\psi)(\cos\Phi_{0}-\cos\phi_{\alpha})\bigg]
\end{split}
\end{align}
and
\begin{align}
B_{\alpha}^{1}(t)&=\frac{1}{\omega_{0}^{1/3}}\sin(2\omega_{0}t)\\
B_{\alpha}^{2}(t)&=\frac{1}{\omega_{0}^{1/3}}\cos(2\omega_{0}t),
\end{align}
where, again, $\omega_{0}$ is the angular orbital frequency of the SMBHB. The log-likelihood ratio is
\be
\begin{split}
\ln\,\Lambda&=\sum_{\alpha=1}^{M}\left[b_{i\alpha}({r_{\alpha}}|B_{\alpha}^{i})-\frac{1}{2}(B_{\alpha}^{i}|B_{\alpha}^{j})b_{i\alpha}b_{j\alpha}\right]\\
&=\sum_{\alpha=1}^{M}\left[b_{i\alpha}{P_{\alpha}}^{i}-\frac{1}{2}Q_{\alpha}^{ij}b_{i\alpha}b_{j\alpha}\right].
\end{split}
\ee
Maximizing the likelihood ratio over the $2M$ 
amplitude parameters $b_{i\alpha}(\zeta,\iota,\psi,\Phi_{0},\phi_{\alpha},\theta,\phi)$ gives
\be
\begin{split}
\frac{\partial \ln\,\Lambda}{\partial b_{k\beta}}&=0=\sum_{\alpha=1}^{M}\bigg[P^{i}_{\alpha}\delta^{k}_{i}\delta_{\alpha\beta} -\frac{1}{2}Q_{\alpha}^{ij}b_{i\alpha}\delta_{j}^{k}\delta_{\alpha\beta}\\
&-\frac{1}{2}Q_{\alpha}^{ij}b_{j\alpha}\delta_{i}^{k}\delta_{\alpha\beta} \bigg]=P^{k}_{\beta}-Q_{\beta}^{ik}b_{i\beta}\\
\end{split}
\ee
which yields the solution for the maximum likelihood estimators of the $2M$ amplitudes
\be
b_{i\beta}=Q_{ik}^{\beta}P_{\beta}^{k}.
\ee
Putting the amplitude estimators back into the likelihood ratio we obtain the $\mathcal{F}_{p}$-statistic
\be
2\mathcal{F}_{p}=\sum_{\alpha=1}^{M}P^{i}_{\alpha}Q^{\alpha}_{ij}P^{j}_{\alpha}.
\ee
It is straightforward to then show that $2\mathcal{F}_{p}$ follows a $\chi^{2}$ distribution with $2M$ degrees of freedom and non-centrality parameter $\hat{\rho}^{2}$ and that
\be
\begin{split}
\langle 2\mathcal{F}_{p}\rangle&=2M+\rho^{2}\\
&=2M+(\tilde{\mathbf{s}}|\tilde{\mathbf{s}})
\end{split}
\ee
where $\rho^{2}=(\tilde{\mathbf{s}}|\tilde{\mathbf{s}})$ is the optimal 
signal-to-noise ratio (see Fig. \ref{fig:pdf}).  Note that this is 
an \emph{incoherent} detection statistic since it involves sum of the 
squares of the data, whereas the Earth-term $\mathcal{F}_{e}$-statistic 
is \emph{coherent} since it involves the square of the sum of the data. 

It is worth pointing out that for the case of white gaussian noise, the 
$\mathcal{F}_{p}$-statistic is  the time domain equivalent to the 
weighted power spectral summing technique studied in \cite{ejm12}. 
For colored gaussian noise the statistic is the time domain 
equivalent to a weighted power spectral summing technique with 
frequency dependent weights. Another feature of this detection 
statistic is that it does not only apply to the low-frequency 
limit. If we work in the high frequency regime where the Earth 
and the pulsar terms are in different frequency bins, we 
can drop the pulsar term and arrive at the exact same maximized 
likelihood function. In this case the pulsar dependence of the 
amplitudes $b_{i\alpha}$ comes from the antenna pattern 
functions the not the pulsar phase. However, many of the 
justifications for dropping the pulsar term mentioned in the 
previous section do not apply in this case since the statistic
is incoherent. We find that this detection statistic will often 
pick out the pulsar term frequency over the Earth term frequency 
because the residuals of Eq. \ref{eq:resids} scale like 
$\omega(t)^{-1/3}$ and the pulsar term will always be at 
an equal or lower frequency than the Earth term frequency 
due to the geometrical delay in Eq. \ref{eq:pTime}. For this 
system of equations we have $2M$ equations and $6+M$ unknowns, 
so if we have 6 or more pulsars we can solve for 
the all the parameters $(\zeta,\iota,\psi,\Phi_{0},\theta,\phi)$ 
along with the pulsar phases $\phi_{\alpha}$.

\subsection{False alarm probability and detection statistics}
\label{sec:prob}
Here we review the false alarm and detection probability distribution 
functions both when the intrinsic parameters are known and unknown. 
Our discussion follows closely that of \cite{jks98} and \cite{jk05}. 
In the case of known extrinsic parameters, we have shown in 
Sections \ref{sec:F} and \ref{sec:Fp} that the statistics 
$2\mathcal{F}$ and $2\mathcal{F}_{p}$ follow $\chi^{2}$ 
distributions with 4 and $4M$ degrees of freedom, respectively, 
when the signal is absent. It was also shown that the aforementioned 
statistics follow a non-central $\chi^{2}$ with non-centrality 
parameters $\bar{\rho}$ and $\rho$, respectively, when the signal is present.

Therefore, the probability distribution functions $p_{0}$ and $p_{1}$ when 
the intrinsic parameters are known and when the signal is absent and present, 
respectively, are
\begin{align}
p_{0}(\mathcal{F})&=\frac{\mathcal{F}^{n/2-1}}{(n/2-1)!}\exp(-\mathcal{F})\\
\begin{split}
p_{1}(\mathcal{F},\kappa)&=\frac{(2\mathcal{F})^{(n/2-1)/2}}
{\kappa^{n/2-1}}I_{n/2-1}\lp\kappa\sqrt{2\mathcal{F}}\rp\\
&\times\exp\lp-\mathcal{F}-\frac{1}{2}\kappa^{2}\rp,
\end{split}
\end{align}
where $n$ is the number of degrees of freedom, $I_{n/2-1}$ is the modified 
Bessel function of the first kind and order $n/2-1$, and $\kappa$ is $\rho$ 
for $\mathcal{F}_{p}$ and $\bar{\rho}$ for $\mathcal{F}_e$. The false alarm 
probability $P_{F}$ is defined as the probability that $\mathcal{F}$ exceeds 
a given threshold $\mathcal{F}_{0}$ when no signal is present. In this case, we have
\be
P_{F}(\mathcal{F}_{0})=\int_{\mathcal{F}_{0}}^{\infty}p_{0}(\mathcal{F})d\mathcal{F}
=\exp(-\mathcal{F}_{0})\sum_{k=0}^{n/2-1}\frac{\mathcal{F}_{0}^{k}}{k!}.
\ee
The probability of detection $P_{D}$ is the probability 
that $\mathcal{F}$ exceeds the threshold $\mathcal{F}_{0}$ when the 
signal-to-noise ratio is $\kappa$:
\be
P_{D}(\mathcal{F}_{0},\kappa)=\int_{\mathcal{F}_{0}}^{\infty}p_{1}(\mathcal{F},\kappa)d\mathcal{F},
\ee
however; we do not deal with the detection probability in this work. 
Our detection criterion is based on the false alarm probability. 

We now turn to the more realistic problem of calculating the false 
alarm probability when the intrinsic parameters are not known. 
A detailed derivation and description is given in \cite{jk00}, 
here we will simply review the result. The probability $P_{F}^{T}$ 
that $\mathcal{F}$ exceeds $\mathcal{F}_{0}$ in one or more cells is given by
\be
\label{eq:fap}
P_{F}^{T}(\mathcal{F}_{0})=1-\left[ 1-P_{F}(\mathcal{F}_{0}) \right]^{N_{c}},
\ee
where $N_{c}$ is the number of independent cells in parameter space. 
The number of independent cells can be calculated via geometrical methods 
described in \cite{jk00} and references therein. 

Here we will make the following approximations. For our $\mathcal{F}_{p}$ 
statistic we will set $N_{p}$ to be equal to the number of independent 
frequency bins defined by the Nyquist frequency. For our $\mathcal{F}_{e}$ 
statistic, we will set $N_{c}$ to be equal to the number of templates used 
in the search. In general the number of independent templates and the 
number of independent cells will be quite different. However,  since we 
only have a three dimensional parameter space and use a nested sampling 
algorithm to conduct the search (thereby reducing the number of 
templates in low likelihood regions of parameter space), setting the 
number of templates equal to the number of independent cells is a reasonable assumption. 

\section{Pipeline, sensitivities, and implementation}
\label{sec:implementation}
In this section we will test the $\mathcal{F}_{e}$ and $\mathcal{F}_{p}$ 
statistics on realistic simulated data sets. First, we will outline our 
detection pipeline, then we will briefly describe our simulated data sets 
and test the ability to confidently detect the signal and recover the 
injected intrinsic parameters. Finally, we perform monte-carlo 
simulations to produce sensitivity curves for PTAs of various configurations
and sensitivities.

\subsection{Detection Pipeline}
\label{sec:pipeline}
The only inputs to our detection pipeline are the ephemeris file (typically called a ``par'' file) and TOA file (typically called a ``tim'' file) for each pulsar. The steps in the pipeline are as follows:
\begin{enumerate}

\item Use the standard pulsar timing package \textsc{Tempo2} \citep{hem06} to form the residuals for each pulsar.

\item Use \textsc{Tempo2} plugin to output the design matrix for each pulsar (see Chapter 15  of \citealt{pvt+92} for more details). Then construct $\mathbf{R}$ from the design matrices following \cite{d07}.

\item Use a maximum likelihood eigenvalue decomposition method described in \cite{exd+12} to make an estimate of $\mathbf{\Sigma}_{\tilde{n}}$. Note that the cross terms in Eq. \ref{eq:cov} are expected to be small, so we will ignore them for this work.

\item Follow the methods described in Secs. \ref{sec:F} and \ref{sec:Fp} to construct the detection statistics and search the relevant parameter space. If using the $\mathcal{F}_{p}$-statistic we simply grid up the frequency space for the search. If using the $\mathcal{F}_{e}$-statistic we use the nested sampling package, MultiNest \citep{fhb09} to search the three dimensional parameter space.

\item Output the maximum value of the detection statistic and number 
of templates used and compute the relevant false alarm probability using Eq. \ref{eq:fap}. 
Here we set our false alarm probability threshold to $10^{-4}$. 
If the false alarm probability corresponding to our maximum 
value of $\mathcal{F}$ is greater than $10^{-4}$ then we claim a detection.

\item Use the maximum likelihood estimators to find the extrinsic parameters 
(using Eqs. \ref{eq:iota}--\ref{eq:zeta}), and
construct the posterior probability distribution to find the 
intrinsic parameters by sampling the maximized likelihood (Eq.~\ref{eq:fstat22}). 
As mentioned above, when using the $\mathcal{F}_{p}$ statistic, 
one could use numerical techniques to obtain estimates of the 
extrinsic parameters.

\item Use the maximum likelihood values of the intrinsic and 
extrinsic parameters to construct Gaussian prior distributions 
and carry out parameter estimation on the the full 7 dimensional 
search space, again using MultiNest, to get better estimates of 
SMBHB parameters.

\end{enumerate}
In this paper we will only conduct steps 1--5 and leave steps 6 and 7 for 
future work. Although this work uses simulated datasets, nothing in this 
detection pipeline makes any assumptions about the spacing of the data, 
or the color of the noise. 

In the absence of a detection we would like to set upper limits on the 
strain amplitude as a function of GW frequency. This can be accomplished
as follows
\begin{enumerate}

\item Run the detection pipeline and determine the value of the $\mathcal{F}$-statistic.

\item For each frequency, choose the value of $\zeta$ corresponding to a specific strain amplitude. 
Then inject a SMBHB signal with randomly drawn binary orientation parameters $(\cos \iota,\psi,\Phi_{0})$.

\item Run the detection pipeline again on this injected data and measure the value of the $\mathcal{F}$-statistic. 

\item Keep the value of $\zeta$ fixed and perform a given number of injections  with different binary orientation parameters (1000, for example) and determine the fraction of $\mathcal{F}$-statistic values that is larger than the value measured in the original data.

\item Repeat steps 2--4 until the strain amplitude is such that 95\% of the injections give a value of the $\mathcal{F}$-statistic that is larger than the original value.

\item Record this value and repeat steps 2--5 at each frequency.

\end{enumerate}

\begin{figure*}[]
  \begin{center}
  \subfigure[]{
  	\includegraphics[width=0.48\textwidth]{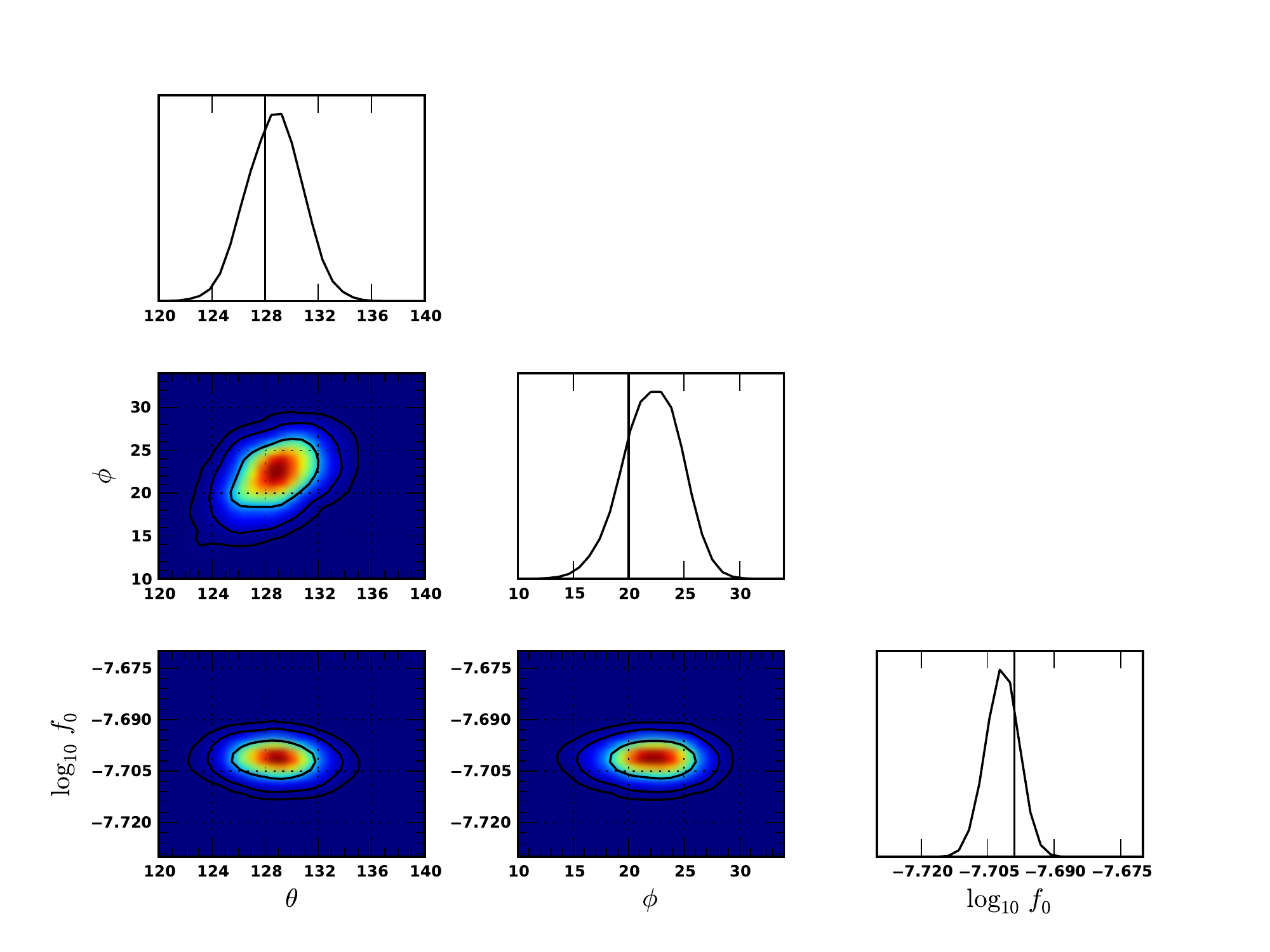}
	\label{fig:feTest_a}}
  \subfigure[]{
  	\includegraphics[width=0.48\textwidth]{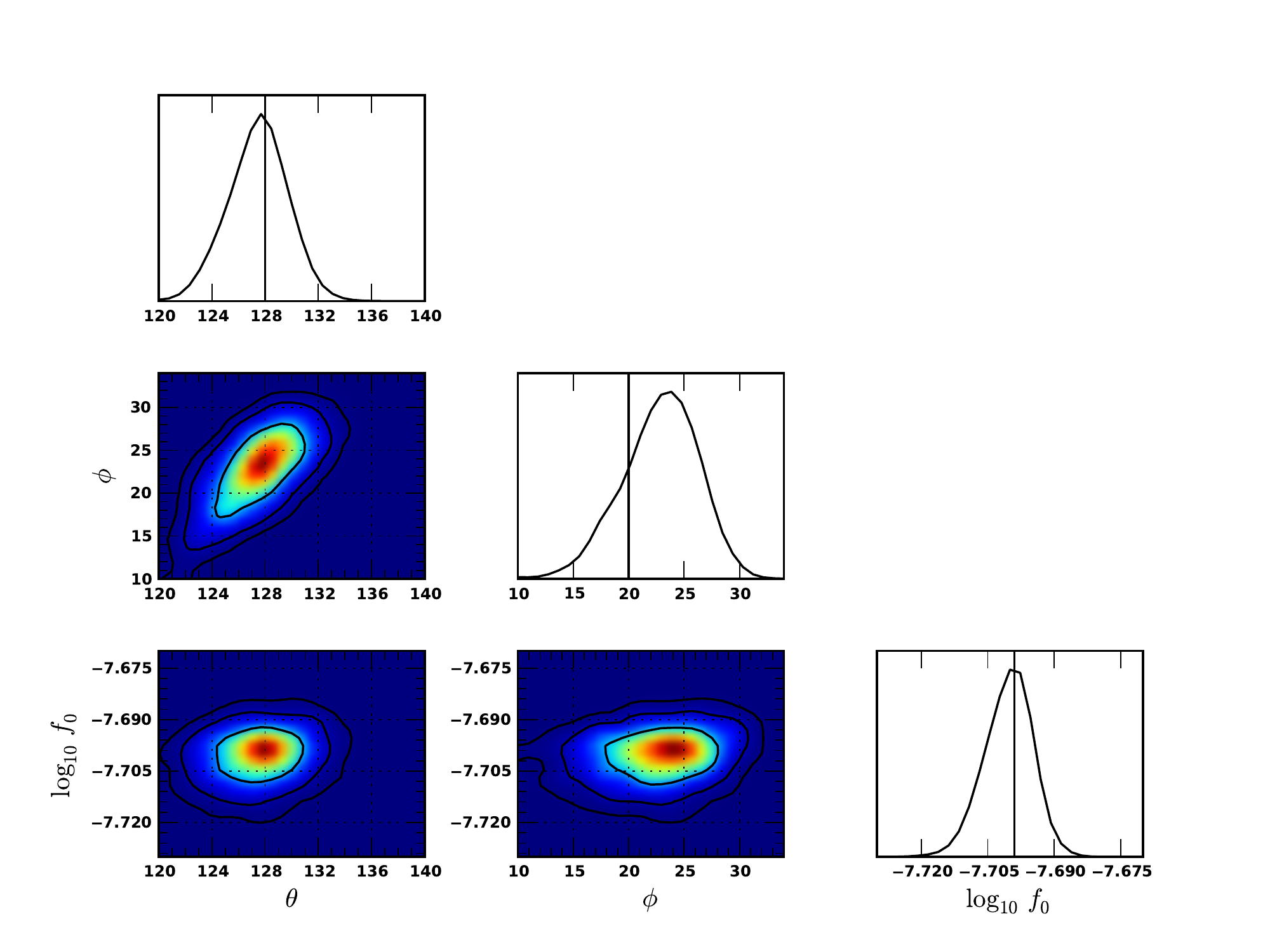}
	\label{fig:feTest_b}}\\
  \end{center}
  \caption{Posterior probability distribution functions for sky location and orbital frequency for a network SNR=14 injection with and without red noise. Here we have used a PTA with 25 pulsars. The vertical lines indicated the injected parameters and the contours are the one, two and three sigma contours. (a): 100 ns white noise. (b): 100 ns white noise and uncorrelated red noise with amplitude $A=4.22\times 10^{-33}{\rm\,s}^{-2.09}$ and $\gamma=4.1$. We see that the sky location and orbital frequency have all been recovered at the one-sigma level in both cases.}
\label{fig:feTest}
\end{figure*}

\subsection{Simulated data sets}
For this work we use a simulated pulsar timing array with sky 
locations drawn from uniform distributions in $\cos\theta$ 
and $\phi$. All pulsars are assumed to have a distance of 
1 kpc and a white noise rms of 100 ns with equal error bars. 
The timespan of the observations for all pulsars is 5 years 
with evenly spaced  bi-monthly TOA measurements. Each set 
of residuals has been created by fitting a full timing model 
including spin-down, astrometric, and binary parameters 
(see \citealt{ehm06} for details). As a check in some 
simulations an uncorrelated red noise process with 
a power law spectrum 
$P(f)=Af^{-\gamma}$ is included in the residuals, which 
has no effect on our results. 
While these simulated data sets do not include uneven 
sampling or extra fitting procedures like jumps or time 
varying DM variations, they do capture the essence of 
real timing residuals in the quadratic fitting of the 
spin-down parameters and the yearly and half yearly 
sinusoidal trends due to the sky location, proper 
motion and parallax fitting. Very uneven sampling 
is likely to reduce our sensitivity at higher 
frequencies and a detailed study of this problem 
will be presented in future work.

\subsection{Implementation of the detection statistics}
Here we will test our detection statistics on mock data sets with injected SMBHB GW signals in the presence of white and red gaussian noise. We will focus primarily on the $\mathcal{F}_{e}$-statistic since, as we will show, it is a more robust detection statistic. Then, we will implement a procedure to produce an upper limit on the GW strain amplitude as a function of frequency for a simulated NANOGrav \citep{dfg+12} array and plausible SKA arrays.

Fig. \ref{fig:feTest} shows the posterior probability distributions of the intrinsic search parameters for simulated SMBHB signals in the presence of 100 ns white noise (Fig. \ref{fig:feTest_a})  and uncorrelated red noise with amplitude $A=4.22\times 10^{-33}{\rm\,s}^{-2.09}$ and $\gamma=4.1$. The two cases do have different realizations of the white noise, however, we can see that the $\mathcal{F}_{e}$-statistic does a very good job of determining the frequency and sky location of the source. In general, the $\mathcal{F}_{e}$-statistic is more robust than the $\mathcal{F}_{p}$ statistic because it produces estimates of the sky location as well as the frequency, which is very important when looking for electromagnetic counterparts. 

It is possible to produce a sensitivity curve by a method that is 
similar to what we use to set upper limits. In this case we use simulated data 
with a given level of noise and no signal present. We follow the 
method presented in Sec. \ref{sec:pipeline} except we now look for 
strain amplitude that gives a false alarm probability that is higher 
than our threshold ($10^{-4}$ in our case) in 95\% of realizations 
for each frequency. For clarity, we define the strain amplitude as
\be
h=2\frac{\mathcal{M}^{5/3}(\pi f_{\rm gw})^{2/3}}{D},
\ee
where $f_{\rm gw}=\omega_{0}/\pi$. This amplitude comes from the 
overall scaling factor that results in differentiating 
Eq. \ref{eq:rplus} and \ref{eq:rcross} with respect to time. 
For simplicity and speed we have simplified this method for our 
sensitivity plots. Instead of performing a search at each 
frequency, we simply evaluate the $\mathcal{F}_e$ and $\mathcal{F}_p$ 
statistics at the values of the  injected parameters. The purpose 
of these sensitivity plots is to illustrate the overall features 
of the different detection statistics and to give order of 
magnitude estimates of expected sensitivity for real data. 
\begin{figure}[b]
  \begin{center}
	\includegraphics[width=0.49\textwidth]{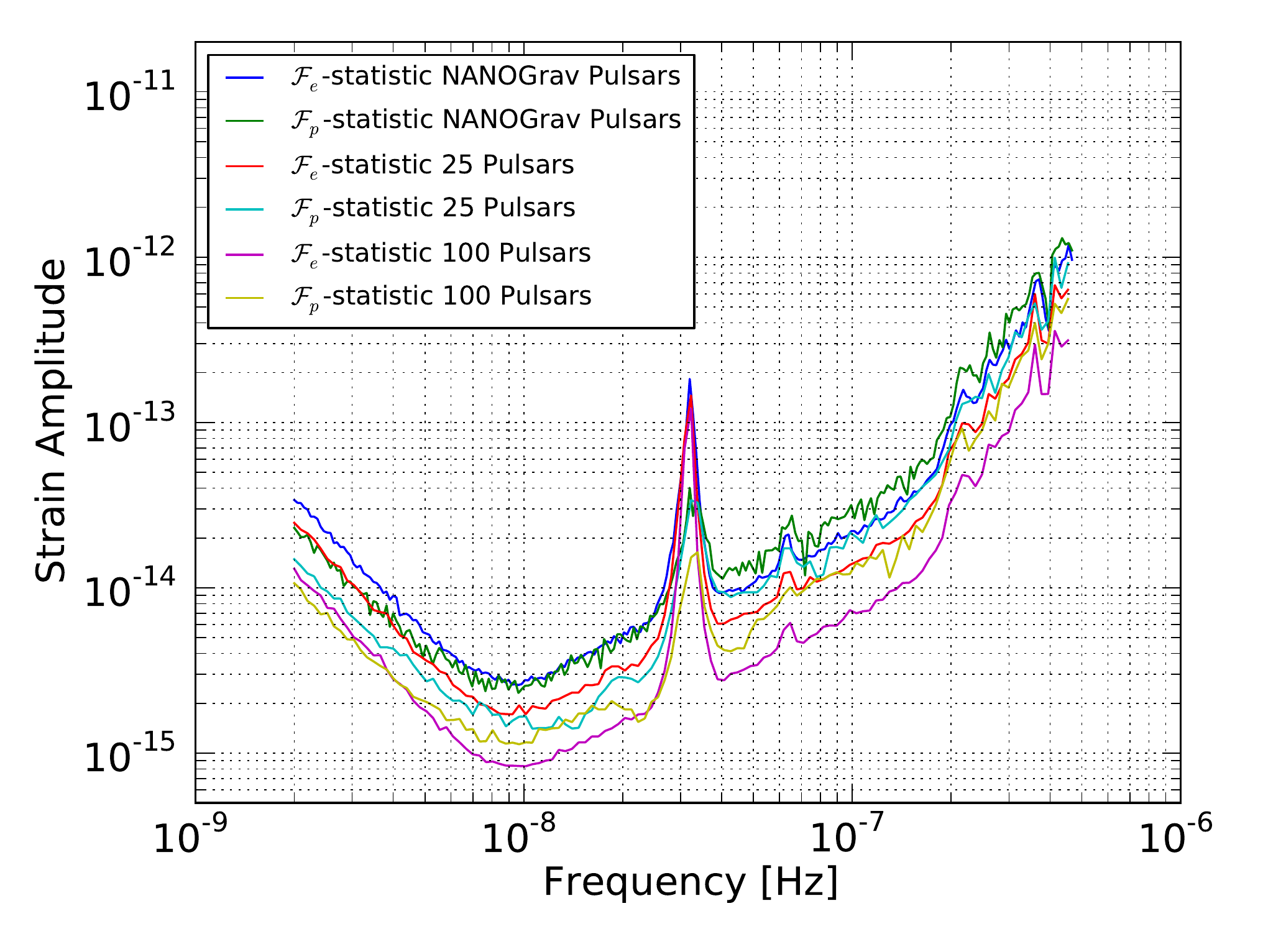}
  \end{center}
  \caption{Sensitivity curves for the $\mathcal{F}_{e}$ and $\mathcal{F}_{p}$ statistics for different PTA configurations (all pulsars have 100 ns residuals). The blue and green lines are the sensitivity curves for the $\mathcal{F}_{e}$ and $\mathcal{F}_{p}$ statistics, respectively, for a simulated NANOGrav PTA. The red and cyan lines are the sensitivity curves for the $\mathcal{F}_{e}$ and $\mathcal{F}_{p}$ statistics, respectively, for a simulated PTA with 25 pulsars. The magenta and yellow lines are the sensitivity curves for the $\mathcal{F}_{e}$ and $\mathcal{F}_{p}$ statistics, respectively, for a simulated PTA with 100 pulsars.}
\label{fig:sens}
\end{figure}

We have produced various sensitivity curves for both the $\mathcal{F}_{e}$ and $\mathcal{F}_{p}$ statistics in Fig. \ref{fig:sens}. The three scenarios that we look at are a 17 pulsar simulated NANOGrav array in which we use the real sky location and timing models of the NANOGrav pulsars, and simulated PTAs with 25 and 100 pulsars at random sky locations. The loss in sensitivity at GW frequencies of $1\,{\rm yr}^{-1}$ and $2\,{\rm yr}^{-1}$ are due to the fitting of the pulsar's sky location and proper motion, and parallax, respectively.   It is important to note that the sensitivity curves for the $\mathcal{F}_{e}$ and $\mathcal{F}_{p}$ statistics in the 17 and 25 pulsar cases, respectively, are very similar. Conversely, for the case of 100 pulsars the $\mathcal{F}_{e}$ statistic is more sensitive by a factor of $\sim 2$ for almost all frequencies. This is due to the different scaling relations of the statistics vs. the number of pulsars ($\mathcal{F}_{e}\propto \sqrt{M}$ while $\mathcal{F}_{p}\propto M^{1/4}$). However, the  plot shows that the $\mathcal{F}_{p}$-statistic is more sensitive at lower frequencies and the $\mathcal{F}_{e}$-statistic is more sensitive at higher frequencies. There are two effects that contribute to this. The first is a result of our simulation and stems from the fact that we assume that for a given frequency, the maximum value of the $\mathcal{F}_{e}$-statistic is at the injected sky location. However, for low frequencies where the Earth and pulsar term are in the same frequency bin this assumption breaks down as the sky location will be biased (see e.g. \citealt{ejm12}). The second effect is one inherent to our detection statistics themselves. As discussed in Sec. \ref{sec:Fp}, the $\mathcal{F}_{p}$-statistic has different meanings in the low and high frequency regimes. In the low frequency regime, it effectively contains the entire signal (Earth and pulsar terms), and in the high frequency regime it only contains the Earth term piece since the pulsar terms are out of that frequency bin. This distinction results in a different scaling relation for the ratio of $\mathcal{F}_{e}/\mathcal{F}_{p}$. In the low frequency case the $\mathcal{F}_{e}$-statistic scales coherently but it only has approximately \emph{half} of the signal, whereas, the $\mathcal{F}_{p}$-statistic scales incoherently but has the \emph{full} signal. Therefore, the ratio scales as $M^{1/4}/2$, thus the incoherent method will do better for $M\le 16$. Conversely, in the high frequency regime, both statistics contain only \emph{half} of the signal and the ratio scales as $M^{1/4}$. Therefore, the coherent statistic will do about a factor of 2 better than the incoherent method for $M \ge 16$. 

\section{Summary and Outlook}
\label{sec:summary}
In this work we have adapted the standard $\mathcal{F}$-statistic \citep{jks98} 
to act as a detection statistic for continuous wave searches in 
realistic PTA data. We have also developed an 
incoherent detection statistic that maximizes over all pulsar contributions 
to the likelihood. Both of these detection statistics are implemented 
in the time domain to avoid spectral leakage problems associated 
with Fourier domain methods applied to irregularly sampled data. These methods 
take the pulsar timing model fitting into account and have been generalized to 
account for both correlated and uncorrelated colored noise. Most of our analysis 
relies on dropping the pulsar term from our signal model as it will not add 
coherently. We have justified the use of this approximation in most 
astrophysically likely scenarios. It was shown that both of the detection 
statistics follow well known $\chi^{2}$ distributions in the presence and 
absence of GW signals and therefore have well defined false-alarm 
probabilities. We have shown that the $\mathcal{F}_{e}$ statistic can 
not only confidently detect a GW signal but can also determine the sky 
location and frequency of the source to relatively high accuracy in the 
presence of white and colored gaussian noise. A realistic implementation 
of a fully functional continuous GW pipeline starting from basic pulsar 
timing data and methods for computing upper limits on the strain amplitude
 were outlined in detail. Finally, we have used simulated data sets of 
various PTA configurations to produce sensitivity curves for our 
$\mathcal{F}$-statistics. From these sensitivity curves, we have shown that the sensitivity of the $\mathcal{F}_{e}$ and $\mathcal{F}_{p}$ statistics are very similar for $M\le25$ pulsars and that the $\mathcal{F}_{e}$ statistic becomes more sensitive for $M>25$ and for higher frequencies.

As was shown in \citet{ejm12}, explicitly searching over the pulsar distances or 
somewhat equivalently, the GW phases at the pulsar locations (in the low frequency 
regime), is computationally prohibitive for $M\gtrsim5$. A statistic that could 
maximize over these GW phases would greatly reduce the parameter space of the 
search, while still preserving the SNR of the full signal. The implementation of 
such an algorithm will be the subject of future work. However, we will give the 
derivation here. From Eq.~\ref{eq:resids}, in the low-frequency limit we can write the 
signal in the following form
\be
s_{\alpha}(t)=\sum_{i=0}^{M}\left[ (\cos\Phi_{\alpha}-1)\delta_{ij}
+\sin\Phi_{\alpha}\varepsilon_{ij} \right]a^{j}A^{i},
\ee
where $\Phi_{\alpha}=\omega L_{\alpha}(1+\omhat\cdot\phat_{\alpha})$, 
$a^{i}=a_{i}$ and $A^{i}$ are defined in Eqs. \ref{eq:amps} and 
\ref{eq:basis}, respectively, and the matix
\be
\varepsilon=\bb 0 & -1 & 0 & 0\\ 1 & 0 & 0 & 0\\ 0 & 0 & 0 & 1\\ 0 & 0 & 1 & 0\\ \eb.
\ee
After some algebra, the log-likelihood ratio of Eq. \ref{eq:llike} can be written as
\be
\label{eq:likePhase}
\begin{split}
\ln\,\Lambda&=\sum_{\alpha=1}^{M}\big[b(\cos^{2}\Phi_{\alpha}
-\sin^{2}\Phi_{\alpha})+c\cos\Phi_{\alpha}\\
&+d\sin\Phi_{\alpha}+f\sin\Phi_{\alpha}\cos\Phi_{\alpha}\big],
\end{split}
\ee
with
\begin{align}
b&=-\frac{1}{2}M_{\alpha}^{ij}a_{i}a_{j}\\
c&=N_{\alpha}^{i}a_{i}+M_{\alpha}^{ij}a_{i}a_{j}\\
d&=N_{\alpha}^{i}\varepsilon_{ij}a^{j}\\
f&=-M_{\alpha}^{ij}\varepsilon_{\ell j}a_{i}a^{\ell}
\end{align}
where $M_{\alpha}$ and $N_{\alpha}$ are defined by the following relations
\begin{align}
M^{ij}_{\alpha}&=(A^{i}_{\alpha}|A^{j}_{\alpha})\\
N^{i}_{\alpha}&=(r_{\alpha}|A^{i}_{\alpha}).
\end{align}
Maximizing the log-likelihood with respect to the pulsar phases $\Phi_{\beta}$, we obtain
\be
\begin{split}
\frac{\partial \ln\,\Lambda}{\partial \Phi_{\beta}}&=f(\cos^{2}\Phi_{\beta}-\sin^{2}\Phi_{\beta})+2b\cos\Phi_{\beta}\sin\Phi_{\beta}\\
&-c\sin\Phi_{\beta}+d\cos\Phi_{\beta}=0.
\end{split}
\ee
Setting $x=\cos\Phi_{\beta}$, this expression reduces to a quartic equation of the form
\be
\label{eq:quart}
\begin{split}
0&=(4f^{2}+16b^{2})x^{4}+(4fd+8cb)x^{3}\\
&+(c^{2}-4f^{2}-16b^{2})x^{2}+(-2fd-8cb)x\\
&+f^{2}-c^{2}
\end{split}
\ee
which is guaranteed to have at least one unique solution.  
This maximization results in a monumental reduction in the parameter space 
that needs to be searched. It takes on the order of $\sim 10^{2M}$ 
templates just to cover the pulsar phases \citep{ejm12}.
In practice, we could 
construct the various quantities $M_{\alpha}$, $N_{\alpha}$, $\mathbf{a}$, $b$, 
$c$, $d$, and $f$, solve Eq. \ref{eq:quart} numerically to find the maximum likelihood 
estimators for all the pulsar phases. 
Substituting these solutions back into our likelihood 
Eq. \ref{eq:likePhase} still leaves us with the problem of searching 
over a 7 dimensional parameter space (since the amplitudes $\mathbf{a}$ depend on 4 parameters 
$(\zeta,\iota,\Phi_{0},\psi)$ and the basis functions $\mathbf{A}$ depend on 
3 parameters $(\theta,\phi,\omega_{0})$). We note, however, that this can be 
easily handled with a Markov chain Monte-Carlo (MCMC) or nested 
sampling algorithm. 

Looking to the future, the pipeline outlined in this paper will be used to 
analyze real pulsar timing residuals and, in the absence of a detection, 
construct upper limits on the strain amplitude as a function of frequency. 
We will also further develop and test our likelihood maximized over the GW 
phase at the pulsar on both simulated and real data. We will also begin to 
generalize the methods discussed in the this paper to deal with eccentric 
signal models.

\acknowledgments

We would like to thank the members of the NANOGrav data analysis working group 
for their comments and support, especially Jim Cordes, Paul Demorest, Rick Jenet, 
Andrea Lommen, Delphine Perrodin, Sam Finn, and Joe Romano. We would also like to thank 
Rutger van Haasteren for developing and making available the \textsc{Tempo2} plugin that 
calculates the design matrices. This work was partially funded by the NSF through CAREER award 
number 0955929, PIRE award number 0968126, and award number 0970074.

\bibliographystyle{apj}
\bibliography{apjjabb,bib}

\end{document}